\documentclass[12pt,a4paper,notitlepage]{article}
\usepackage{amsmath}
\usepackage{amssymb}
\usepackage{graphicx}
\usepackage{hyperref}
\usepackage{times}
\usepackage{epstopdf}

\begin{document}

\author{Yi Wang, Yujie Chen, Yanfeng Zhang$^*$, Hui Chen and Siyuan Yu\\
State Key Laboratory of Optoelectronic Materials and Technologies,\\ 
School of Electronics and Information Technology, \\
Sun Yat-sen University, Guangzhou 510275, China
\\ $^*$Corresponding author: zhangyf33@mail.sysu.edu.cn}

\title{Generalised Hermite-Gaussian beams and mode transformations}

\maketitle 

\begin{abstract}
Generalised Hermite-Gaussian modes (gHG modes), an extended notion of Hermite-Gaussian modes (HG modes), are formed by the summation of normal HG modes with a characteristic function $\alpha$, which can be used to unite conventional HG modes and Laguerre-Gaussian modes (LG modes). 
An infinite number of normalised orthogonal modes can thus be obtained by modulation of the function $\alpha$. 
The gHG mode notion provides a useful tool in analysis of the deformation and transformation phenomena occurring in propagation of HG and LG modes with astigmatic perturbation.
\end{abstract}

\section{Introduction}

Light fields with irreducible point dislocations \cite{nye1974dislocations} and hence optical vortices (OV) have drawn much attention since the 1990s \cite{allen1992orbital, beijersbergen1993astigmatic}, for their broad scope of applications and intriguing physical nature. 
Potential applications include optical trapping, optical communications and imaging. 
Among optical modes with OVs, LG modes, those originate from cylindrical resonant cavities, are studied most intensively in recent years. 
On the other hand, HG modes, one of the most familiar mode-sets \cite{siegman1986lasers}, has long been studied since discovery, yet new concepts about it have continued to emerge \cite{Abramochkin:2004fq, Abramochkin2010general, dong2013transformations, Kotlyar:2014fp}. 
In 1991, HG and LG modes were linked by Abramochkin \emph{et al.} via the production of LG beams from astigmatic-lens-twisted HG beams \cite{Abramochkin:1991ff}. 
LG modes can be obtained by passing HG modes through properly arranged cylindrical lens pairs. 
Later this idea was improved into a united notion called Hermite-Laguerre-Gaussian beams (HLG beams) along with a beam parameter $\alpha$, in which LG beams and HG beams are merely two special cases \cite{Abramochkin:2004fq}. 
A continuous evolution between HG and LG modes can be achieved by adjusting the astigmatic parameter $\alpha$ of the lens. 
The modes in the evolution process sitting between LG and HG modes have split OVs. 
In 2007, a new kind of canonical transform, gyrator transform (GT) \cite{Rodrigo:2007gyrator, Rodrigo:2007exper}, was introduced to the optics community, further generalising the relation between HG and LG beams, by regarding LG beams as the result of a gyrator transform of HG beams with the transform angle $\alpha=\pi/4$. 
Various OV-split states are obtained by changing the GT angle $\alpha$. 
This work also extends the relation between HG and LG modes into relations between many other optical modes with edge dislocations and point dislocations.

Here, we propose a new concept called generalised HG (gHG) modes. 
gHG modes are obtained by superposition of phase-delayed HG modes. 
With this new notion, the OV split phenomena of LG beams under astigmatic transformation \cite{Bekshaev:2003hb, Bekshaev:2004ez, Bekshaev:2008as} can be reproduced and analysed. 
gHG modes also lead us into uncovering an infinite number of orthogonal optical mode types, including modes resembling the formerly defined HLG modes, with a phase delay.

\section{Generalised HG beams}

\subsection{Mode Definition}
Assuming $N=n+m$, LG beams can be written in the form of superposition of HG beams with a $\pi/2$ phase delay \cite{beijersbergen1993astigmatic}:
\begin{equation}\label{LG}
U^{\text{LG}}_{nm}\left(x,y,z\right)=\sum^{N}_{k=0}i^kb\left(n,m,k\right)U^{\text{HG}}_{N-k,k}\left(x,y,z\right),
\end{equation}
where
\begin{equation}
b\left(n,m,k\right)=\sqrt{\frac{\left(N-k\right)!k!}{2^Nn!m!}}\frac{1}{k!}\frac{d^k}{dt^k}\left[\left(1-t\right)^n\left(1+t\right)^m\right]_{t=0},
\end{equation}
\begin{equation}
\begin{aligned}
U^{\rm{HG}}_{nm}(x,y,z) &=C \frac{1}{w} \exp\left(-\frac{x^2+y^2}{w^2}\right) \mathcal{H}_n\left(\frac{\sqrt{2}x}{w}\right)\mathcal{H}_m\left(\frac{\sqrt{2}y}{w}\right) \\
&\quad\times \exp\left[-i\frac{k(x^2+y^2)}{2R}\right]\exp\left[-i(n+m+1)\Psi\right]
\end{aligned}
\end{equation}
with
\begin{equation}
\left\{
\begin{aligned}
&C = \sqrt{\frac{2}{\pi n! m!}}2^{-\frac{N}{2}}\\
&z_R = \frac{1}{2}kw^2(0)\\
&R(z) = \frac{z^2+z^2_R}{z}\\
&\frac{1}{2}kw^2(z)=\frac{z^2+z^2_R}{z_R}\\
&\Psi(z)=\arctan\frac{z}{z_R}\\
\end{aligned} \right. .
\end{equation}

Modifying the phase factor $i$ in Eq. \eqref{LG} into an arbitrary one gives the definition of gHG modes:
\begin{equation}\label{def}
U_{nm\alpha(k) }^{g{\rm{HG}}} = \sum\limits_{k = 0}^N {\exp } [i\alpha(k) ]b(n,m,k)U_{N - k,k}^{{\rm{HG}}},
\end{equation}
where the characteristic function $\alpha(k)$ is an arbitrary real function of the summation parameter $k$.

\begin{figure}\includegraphics[width = 14cm]{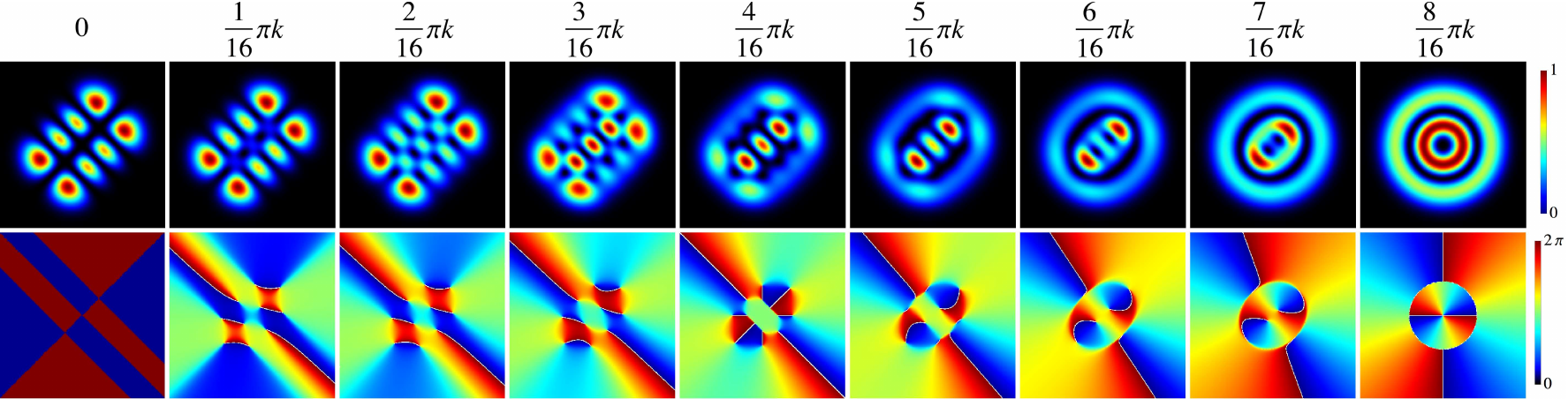}
\centering
\caption{Numerical simulation of the transverse intensity distributions (upper row) and phase distributions (lower row) of mode $U_{13,\alpha}^{g\rm{HG}}(x,y)$ with $\alpha(k)=Ck$, where the coefficient $C$ varies from 0 to $\pi/2$. 
Colour bars are applied to each row. 
The intensity and phase distributions transform from $U_{13}^{\rm{HG}}$ to $U_{13}^{\rm{LG}}$ according to $C$.}
\label{wide}
\end{figure}

\subsection{Mode Characterisation}
For gHG modes, every particular $\alpha(k)$ produces a unique child modes set. 
Examples of these child modes sets are LG modes with $\alpha(k) = q(\pi/2)k$, in which $q$ being an arbitrary odd integer and HG modes (diagonal) with $\alpha(k) = q(\pi/2)k$, with $q$ being an arbitrary even integer. 
When $\alpha(k)$ is any other linear real functions of $k$, gHG modes collapse into HLG-like modes. 
Variation of the coefficient $C$ in $\alpha(k)=Ck$ gives rise to an evolution between HG beams and LG beams, as is seen in Fig. \ref{wide}. 
New mode-sets appear if $\alpha(k)$ takes the form of other real functions. 
On account of the arbitrariness of $\alpha(k)$, the number of mode-sets in the gHG family is virtually infinite. 

\begin{figure}[!ht]
\centering
\includegraphics[width = 11cm]{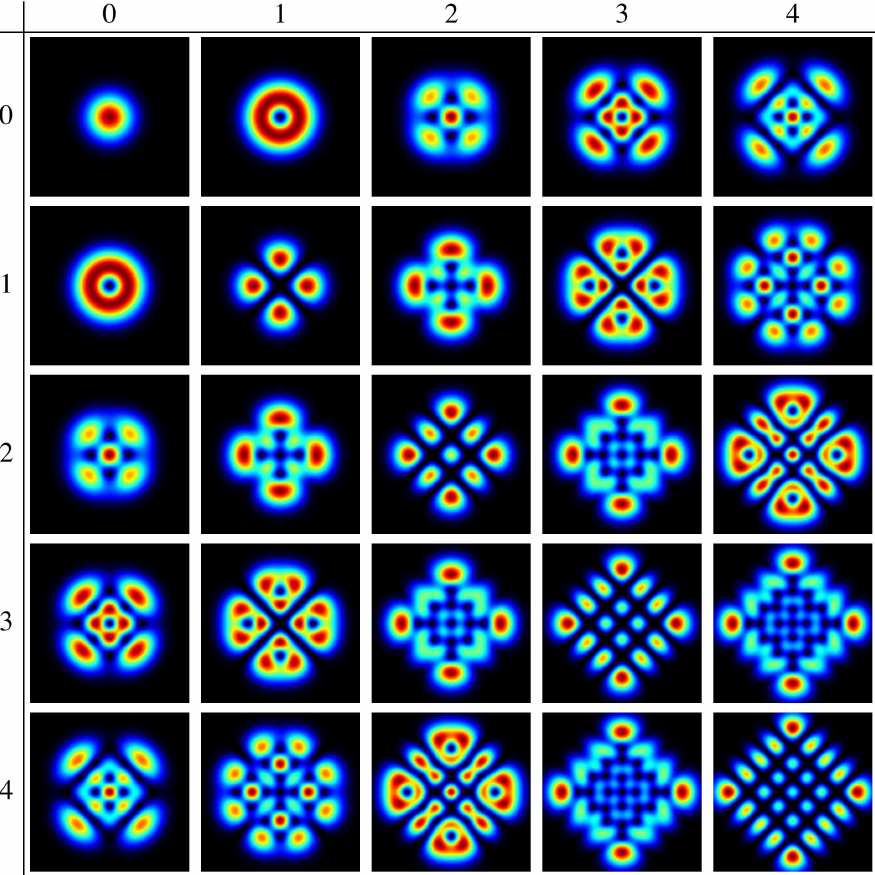}
\caption{Numerical simulation of the transverse intensity distributions gHG modes with $\alpha(k)=(\pi/2)k^2$. 
Labeled columns represent the modes with index $n$, meanwhile labeled rows represent the modes with index $m$. 
For $n=m$ the gHG modes devolve into HG modes, for $nm=01~or~nm=10$ the gHG modes devolve into LG modes. 
Note that modes with $n_1=m_2,~m_1=n_2$ have opposite topological charges, while the intensity distribution remains the same.}
\label{gHG}
\end{figure}

Note that several important beam features, such as propagation stability, orthogonality and normalisation, can be derived directly from Eq. \eqref{def} without knowing the explicit expressions. 
Every mode in the gHG modes set is propagation stable, since its building parts (HG bases) have constant phase differences along propagation. 
We also note that the whole gHG mode-set is over-complete. 
The transverse intensity distributions of gHG modes with a particular instance of $\alpha(k)$ are shown in Fig. \ref{gHG}.

To demonstrate the orthogonality and normalisation of gHG modes of a given $\alpha(k)$, one can perform the following calculation:
\begin{equation}\label{ortho}
\begin{aligned}
&\iint\limits_{{\mathbb{R}^2}} U_{n_1m_1\alpha }^{g{\rm{HG}}}(x,y)\overline {U_{n_2m_2\alpha }^{g{\rm{HG}}}(x,y)} dxdy \\
&=\iint\limits_{{\mathbb{R}^2}} \sum\limits_{k_1 = 0}^{N_1} \exp\left[i\alpha(k_1)\right]b(n_1,m_1,k_1)U_{N_1 - k_1,k_1}^{{\rm{HG}}}\\
&\qquad\times \sum\limits_{k_2 = 0}^{N_2} \exp\left[-i\alpha(k_2)\right]b(n_2,m_2,k_2)\overline{U_{N_2 - k_2,k_2}^{{\rm{HG}}}} dxdy ,
\end{aligned} 
\end{equation}
where the overbar represents a complex conjugate. 
With the help of the orthogonality relations of Hermite polynomials
\begin{equation}
\int_{ - \infty }^\infty  {{{\cal H}_n}(x)} {{\cal H}_m}(x)\exp \left( { - {x^2}} \right)dx = \sqrt \pi  {2^n}n!{\delta _{nm}} .
\end{equation}
we can find that, only the terms with $k_1=k_2$ and $N_1-k_1=N_2-k_2$ give non-zero results. Rewriting both $k_1$, $k_2$ into $k$ and both $N_1$, $N_2$ into $N$, Eq. \eqref{ortho} can be reduced into the following expression independent of function $\alpha(k)$:
\begin{equation}\label{ortho2}
\iint\limits_{{\mathbb{R}^2}} \sum\limits_{k = 0}^{N}b(n_1,m_1,k)U_{N - k,k}^{{\rm{HG}}} b(n_2,m_2,k)\overline{U_{N - k,k}^{{\rm{HG}}}} dxdy .
\end{equation}
By comparing Eq. \eqref{ortho2} with the orthogonality relations of LG modes, the orthogonality and normalisation relations of gHG modes can be derived:
\begin{equation}
\begin{aligned}
\iint\limits_{{\mathbb{R}^2}} U_{n_1m_1\alpha }^{g{\rm{HG}}}(x,y)\overline {U_{n_2m_2\alpha }^{g{\rm{HG}}}(x,y)} dxdy &= \iint\limits_{{\mathbb{R}^2}} U_{n_1,m_1}^{{\rm{LG}}}\overline{U_{n_2,m_2}^{{\rm{LG}}}}dxdy \\
&= {\delta _{{n_1}{n_2}}}{\delta _{{m_1}{m_2}}} .
\end{aligned}
\end{equation}

Under certain circumstances, the summation expression in Eq. \eqref{def} can be expressed explicitly. 
When $\min(n,m)=0$ and $\alpha(k)=Ck$ with $C$ being a constant, assume $m=0$ and $n=\ell$, Eq. \eqref{def} can be simplified using the summation theorem of Hermite polynomials \cite{jahnke1960tables} as follows:
\begin{equation}\label{p=0}
\begin{aligned}
U^{g\textrm{HG}}_{\ell 0 (Ck)}&=\sum_{k=0}^{\ell} \text{e}^{iCk} b(\ell,0,k) U^{\textrm{HG}}_{\ell-k,k}\\
&\propto \exp \left( { - \frac{{{x^2} + {y^2}}}{{{w^2}}}} \right)\sum_{k=0}^{\ell} \frac{\left(\text{-e}^{iC}\right)^k}{(\ell-k)!k!} {\cal H}_{\ell-k}\left(\frac{\sqrt{2}x}{w}\right){\cal H}_{k}\left(\frac{\sqrt{2}y}{w}\right)\\
&\propto \exp \left( { - \frac{{{x^2} + {y^2}}}{{{w^2}}}} \right){\cal H}_\ell \left[ \left(\frac{\sqrt{2}x}{w}-\text{e}^{iC}\frac{\sqrt{2}y}{w}\right)\times\left(\text{e}^{i2C}+1\right)^{-\frac{1}{2}}\right],
\end{aligned}
\end{equation}
where the plane $(x,y,0)$ is implied.
Furthermore, the coordinates of embedded OVs can be readily obtained:
\begin{equation}
x=\textrm{sgn}(\cos C)y= r_{k,\ell}\sqrt{\frac{w^2}{2-2|\sec C|}},
\end{equation}
where $r_{k,\ell}$ stands for the $k$th root of the $\ell$-order Hermite polynomial.

\section{Mode transformation induced by astigmatic perturbation}
It is known that HG and LG modes undergo significant deformation and distortion when going through astigmatic lenses \cite{Abramochkin:1991ff,vaity2013measuring, Vladimir2010determin}. 
For instance, when refracted by a plane dielectric surface, high order OVs embedded in LG beams tend to split into a set of single-order OVs placed in a row. 
With the notion of gHG modes, this distortion can be regarded as a transformation between modes with the same mode index $nm$ but with different characteristic function $\alpha(k)$, if trivial effects such as the variance of Fresnel refraction coefficients \cite{Sasada:2008bq} and asymmetry in axis $X$ and $Y$, are ignored. 

For simplicity, let us consider the situation of the refraction of a LG beam with $\min(n,m)=0$, which corresponds to a gHG mode with $\min(n,m)=0$ and $\alpha(k)=(\pi/2)k$, by a plane interface. 
Since the incident beam is virtually a summation of HG beams, we can treat the refraction of the LG beams by analysing the behaviour of each HG base. 
Refracted HG beams are
\begin{equation}
\begin{aligned}
U_{nm}^{\textrm{HG}}(x,y,z) \propto & C_1 \exp \left[ { - \left( {\frac{{{x^2}}}{{{w_x}^2}} + \frac{{{y^2}}}{{{w_y}^2}}} \right)} - \frac{{i2\pi n_1}}{\lambda }\left( {\frac{{{x^2}}}{{2{R_x}}} + \frac{{{y^2}}}{{2{R_y}}}} \right) \right] \\
&\times \exp \Bigg\{  - i\Bigg[ \left( {n + \frac{1}{2}} \right){\arctan}\left( {\frac{z}{{{z_{rx}}}}} \right)+ \left( {m + \frac{1}{2}} \right)\\
&\qquad\quad~~~\times{\arctan}\left( {\frac{z}{{{z_{ry}}}}} \right) \Bigg] \Bigg\}{\mathcal{H}_n}\left( {\frac{{\sqrt 2 x}}{{{w_x}}}} \right){\mathcal{H}_m}\left( {\frac{{\sqrt 2 y}}{{{w_y}}}} \right) ,
\end{aligned}
\end{equation}
where $C_1,\lambda,n_1,R_i,w_i,z_{ri}(i=x,y)$ are the normalisation constant, the wavelength in vacuum, the relative refractive index of the medium holding the refracted beam against  the one holding the incident beam, the radii of curvature of wave fronts, the beam transverse sizes in orthogonal directions and the Rayleigh lengths for the axes $X$ and $Y$, respectively. 
By adding up the corresponding HG beams using Eq. \eqref{LG}, the refracted beams can be obtained. 
Simplification and consolidation of terms should lead to figuring out the overall phase differences between each adjacent HG bases ($U^{\textrm{HG}}_{N-k,k}$ and $U^{\textrm{HG}}_{N-k+1,k-1}$):
\begin{equation}\label{cdalpha}
\Delta\Phi_{k,k-1}(z)=\arctan\frac{z}{z_{rx}}-\arctan\frac{z}{z_{ry}}+\frac{\pi}{2} ,
\end{equation}
which corresponds to the coefficeint $C$ in the characteristic function $\alpha(k)=Ck$. 
Note that the phase difference $\Delta\Phi$ is essentially introduced by the astigmatic Gouy phase caused by refraction. 
Rewriting Eq. \eqref{cdalpha} explicitly with (we can always do so by a proper establishment of the coordinate system) $z_0=z/z_R=z/z_{ry}$ and $\theta$ being the incident angle leads us to:
\begin{equation}\label{characterfunction}
C(z_0)=\arctan\left(\frac{z_0 \cos ^2\theta}{1-n_1^2\sin^2\theta}\right)-\arctan z_0+\frac{\pi }{2} .
\end{equation}

\begin{figure}
\centering
\includegraphics[width = 12cm]{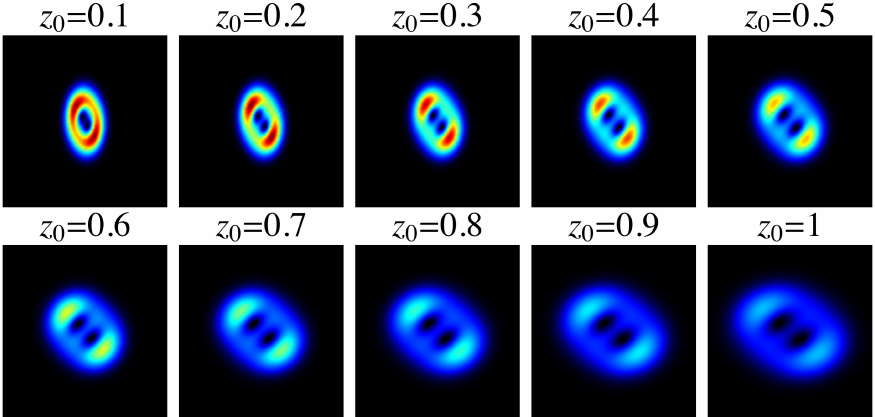}
\caption{Numerical simulation of the transverse distributions of a $U^{\textrm{LG}}_{02}$ mode after refraction at an interface with an incident angle $\theta=\pi/5$ and relative refractive index $n_1=2/3$ (from glass to air). 
The intensity distributions are simulated at different propagation lengths. 
Blue colour indicates lower intensities while red colour indicates higher intensities.}
\label{refraction}
\end{figure}

\begin{figure}
\centering
\includegraphics[width = 12cm]{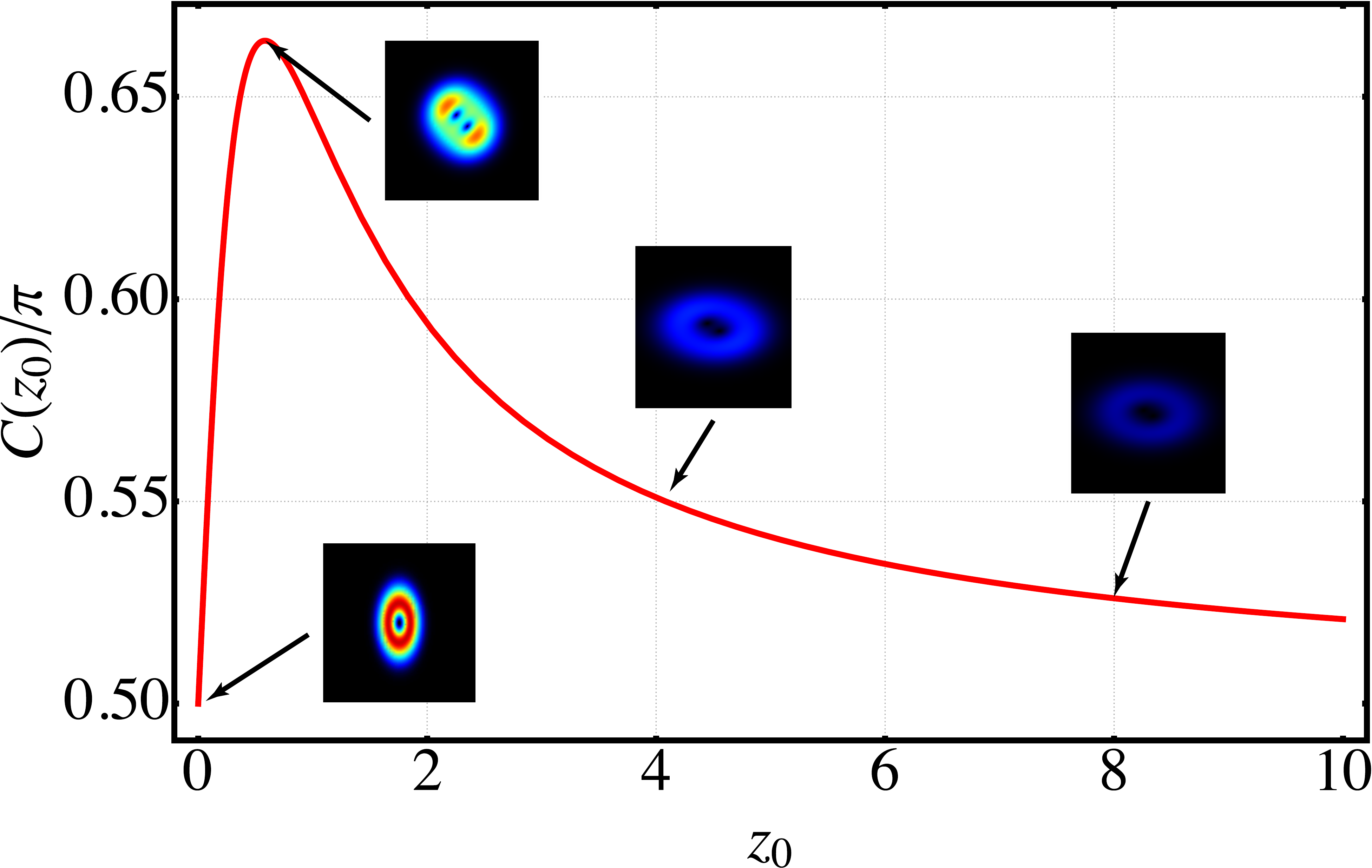}
\caption{Function plot of $C(z_0)/\pi=\Delta\Phi(z_0)/\pi$ with $n_1=2/3$ and the incident angle $\theta = \pi/5$. 
A maximum is expected at $z_0\rightarrow0.58$. 
Correspondingly marked are the simulated amplitude distributions using normalised coordinates in the cross-section.}
\label{cdalpha_f}
\end{figure}

We see here the characteristic function $\alpha(k)=C(z_0)k$ varies with respect to the propagation distance $z_0$, meaning the transverse patterns of the transmitted beams are transforming continuously, as is seen in Fig. \ref{refraction}. 
The rotation of the transverse shape of the refracted beam is driven by astigmatism. 
Apart from scaling and rotation, split OVs also appear in the refracted field, meanwhile the beam tends to transform into a diagonal HG-like mode. 
This transformation is essentially caused by the evolution of $C(z_0)$.
The more $C(z_0)$ deviates from its original value $\pi/2$, which stands for a LG mode, towards $0$ or $\pi$, which stands for a HG mode, the more the beam shape is deformed, as is shown in Fig. \ref{cdalpha_f}. 

The maximum in $C(z_0)$ occurs when $C'(z_0)=0$. 
Solving the equation gives
\begin{equation}
z_0=\sqrt{\frac{1-n_1^2\sin^2\theta}{\cos^2\theta}} ,
\end{equation}
which is equivalent to the most symmetric position, aka the position when condition $w_x=w_y$ is met. 
Note that the position of this maximum depends only on the astigmatic parameters such as $n_1$ and $\theta$, suggesting every $C(z_0)$ of these modes achieve maxima simultaneously regardless of their wavelengths and mode indices. 
Such a feature enables us to detect topological charges of OVs embedded in beams with different wavelengths and mode indices using refraction, only by knowing the Rayleigh ranges of the incident beams. 
When $z_0\rightarrow\infty$, function $C(z_0)\rightarrow\pi/2$, indicating that the deformed modes tend to recover their original state after long distance of propagation. 
This analysis is also reasonable for high order modes, since the Gouy phase changes linearly depending on mode indices.

\section{Conclusion}
In summary, a new notion of gHG modes, with a new characteristic function $\alpha(k)$, is investigated. 
Taking advantage of this notion, conventional HG modes and LG modes are united and inter-transformable via adjusting the beam parameter $\alpha(k)$. 
Intermediate states between pure HG and LG modes, which often occur with the existence of astigmatic perturbations, are also contained in this new concept. 
Using the notion of gHG modes opens an new path to analyse mode transformation phenomena, which may also help dealing with particular crosstalk types in mode division multiplexing systems, such as orbital angular momentum (OAM) communication systems. 
In addition, by designing the characteristic function $\alpha(k)$, one can readily obtain an infinite number of stable orthogonal mode sets in free space, providing more freedom in high speed optical communications.

\section*{Acknowledgement}
This work is supported by the National Basic Research Program of China (973 Program) (No. 2014CB340000), and the Natural Science Foundations of China (No. 61490715). 
Y. Chen would like to thank the Specialised Research Fund for the Doctoral Program of Higher Education of China (No. 20130171120012), and the Fundamental Research Funds for the Central Universities of China (Sun Yat-sen University, No. 15lgpy04). 
Y. Wang thanks the undergraduate training program of the National Physics Base, under the project number 2013014. Thank Huazhou Chen and Shimao Li for fruitful discussion.


\end{document}